\documentclass[aps, showpacs, preprintnumbers]{revtex4}
\usepackage{epsfig}
\begin{document}

\preprint{IMSc/02/03/06}
\title{Quantum Gravity on $dS_{3}$}
\author{  T. R. Govindarajan$^1$\footnote{e-mail : trg@imsc.ernet.in},  
R. K. Kaul$^1$\footnote{e-mail : kaul@imsc.ernet.in}
and V. Suneeta$^2$\footnote{e-mail: suneeta@theorie.physik.uni-muenchen.de}}
			
\affiliation{ $^1$ The Institute of Mathematical Sciences,
CIT Campus, Chennai 600113, India\\
$^2$ Theoretische Physik, Ludwig-Maximilians Universit\"{a}t, \\
Theresienstrasse 37,
D-80333, M\"{u}nchen, Germany \\}
			       
\begin{abstract}
We study quantum gravity on $dS_{3}$ using the Chern-Simons formulation of three-dimensional
gravity.
We derive an exact expression for the partition function for quantum gravity on
$dS_{3}$ in a
Euclidean path integral approach. We show that the topology of the space relevant for 
studying de Sitter entropy is a solid torus.
The quantum fluctuations of de Sitter space are sectors of
configurations of point masses taking a
{\em discrete} set of values. The partition function gives the correct
semi-classical entropy. The
sub-leading correction to the entropy is logarithmic in horizon area,
with a coefficient $-1$.
We discuss this correction in detail, and show
that the sub-leading correction to the entropy
from the dS/CFT correspondence agrees with our result.
A comparison with the corresponding results for the $AdS_3$ BTZ black hole is
also presented. 
\end{abstract}
\maketitle
\section{Introduction}
It is well-known that for a timelike observer in de Sitter space, there exists a horizon, and
regions of spacetime beyond it are not accessible to this observer.
The thermodynamics of such a cosmological
horizon is very similar to that of black hole horizons \cite{gh}. In particular, there is an entropy
associated with the de Sitter spacetime, which is a measure of the information loss (for the timelike
observer) across the cosmological horizon. There have been several
attempts to describe this entropy - for example, 
in terms of microscopic degrees of freedom at the cosmological 
horizon \cite{ms,flin} or in a Chern-Simons
formulation \cite{banados, park}. In fact, in \cite{park}, the entropy of
a general Kerr-de Sitter space is computed. 
More recently, de Sitter entropy has been computed using a CFT at past or future 
infinity \cite{vb, myung, kabat} motivated by the dS/CFT
correspondence \cite{strominger, klemm}. It is shown in \cite{leshouches} 
that the asymptotic symmetry group
of quantum gravity in $dS_{3}$ is the Euclidean conformal group in two dimensions. This has led to the
proposal of a correspondence between quantum gravity in $dS_{3}$ and a CFT at asymptotic infinity. In
\cite{vb}, the entropy of $dS_{3}$ is computed by applying Cardy's formula for the growth of states
in the asymptotic CFT - but as discussed there, there are many subtleties involved. Further, it is not
clear how the entropy associated with information loss across the horizon is described by states
at asymptotic infinity.

We study quantum gravity on $dS_{3}$ using 
the Chern-Simons formulation of Euclidean gravity in three
dimensions which is described in section 3. 
Euclidean $dS_{3}$ has the topology of a three-sphere - we 
show, however, that the topology of the space that is relevant for 
studying the degrees of freedom that contribute to entropy is that of a
solid torus. The degrees of freedom could be thought of as 
living on the boundary torus.
Such a picture of entropy arising from degrees of freedom associated with
the boundary has been studied earlier
in the context of black holes in several approaches,
for example \cite{ponzano, bal, carl, gsv, gks}.
Three dimensional Euclidean gravity 
with a positive cosmological
constant can be described by two $SU(2)$ Chern-Simons theories 
\cite{achtow, witten}. Then, 
$SU(2)$  Wess-Zumino conformal field  theories are naturally induced 
on the boundary \cite{witten1}. The quantum degrees of freedom corresponding
to the de Sitter entropy are described by these conformal field theories. 
In Section 4, we derive an {\em exact} expression for 
the canonical partition
function for $dS_{3}$ in a Euclidean path integral approach.
Considerations of gauge invariance necessitate a 
{\em discrete} sum in the partition
function over point mass configurations as well, upto a certain maximum
value of the mass. The significance of this discrete sum in the partition
function will be discussed in Section 5.
The partition function gives the correct semi-classical entropy for
de Sitter space. The next-order correction to the entropy is 
logarithmic in the horizon "area" \cite{majumdar}.
In the last section, we comment on the logarithmic correction
to the semi-classical entropy. We compare the coefficient of
this correction with that obtained using the dS/CFT correspondence
and find that they agree. We also make a detailed comparison
with similar logarithmic terms in the entropy of the BTZ 
black hole. The comparison suggests a connection between the
regime considered in the black hole parameter space, and
the coefficient of the logarithmic correction.

\section{de Sitter gravity as a Chern-Simons theory}
The gravity action $I_{grav}$ in three dimensions
written in a first-order formalism (using triads $e$ and spin connection 
$\omega$) is the difference of two Chern-Simons actions.
For Lorentzian gravity with a positive cosmological constant, 
\begin{eqnarray}  
I_{\hbox{\scriptsize grav}}
 ~ = ~I_{\hbox{\scriptsize CS}}[A]~ - ~I_{\hbox{\scriptsize CS}}[\bar A] ,
\label{b9}
\end{eqnarray}
where
\begin{eqnarray}
A ~=~ \left(\omega^a + \frac{i}{l}~ e^a\right) T_a , \qquad
\bar A ~=~ \left(\omega^a - \frac{i}{l}~ e^a\right) T_a
\label{b8}
\end{eqnarray}
are  $\hbox{SL}(2,{\bf C})$ gauge fields (with $T_a=-i\sigma_a/2$).
Here, the positive cosmological constant $\Lambda ~=~  (1/{l^2})$.
The Chern-Simons action $I_{\hbox{\scriptsize CS}}[A]$ is
\begin{eqnarray}
I_{\hbox{\scriptsize CS}} ~=~ {k\over4\pi}\int_M
\hbox{Tr}\left( A\wedge dA + {2\over3}A\wedge A\wedge A \right)
\label{cs}
\end{eqnarray}
and the Chern-Simons coupling constant is
$k = -l/4G$.
 
Now, for a manifold with boundary, the Chern-Simons field theory is
described by a Wess-Zumino conformal field theory on the boundary.
Under the decomposition
\begin{eqnarray}
A ~=~ g^{-1}dg ~+~ g^{-1}\tilde A g~ ,
\label{b2}
\end{eqnarray}
the Chern-Simons action (\ref{cs}) becomes \cite{elit}, \cite{ogura}
\begin{eqnarray}
I_{\hbox{\scriptsize CS}}[A]
~  = ~I_{\hbox{\scriptsize CS}}[\tilde A]
~+~ k I^+_{\hbox{\scriptsize WZW}}[g,\tilde A_z] ~,
\label{b3}
\end{eqnarray}
where $I^+_{\hbox{\scriptsize WZW}}[g,\tilde A_z]$ is the action of
a chiral $SU(2)$ Wess-Zumino model on the boundary $\partial M$,
\begin{eqnarray}
I^+_{\hbox{\scriptsize WZW}}[g,\tilde A_z]
&=& \frac{1}{4\pi}\int_{\partial M}\hbox{Tr}
\left(g^{-1}\partial_z g\,g^{-1}\partial_{\bar z} g
~-~ 2g^{-1}\partial_{\bar z} g {\tilde A}_z\right) \nonumber \\
&+& \frac{1}{12\pi}\int_M\hbox{Tr}\left(g^{-1}dg\right)^3 .
\label{b4}
\end{eqnarray}
The `pure gauge' degrees of freedom $g$ are now true
dynamical degrees of freedom at the boundary.
 
We are interested in quantum gravity on $dS_{3}$.
Global $(2+1)-d$ de Sitter spacetime  is described by the metric
\begin{eqnarray}
ds^2 ~=~ - l^2 d\tau^{2} + l^2 \cosh^{2} \tau d\Omega^2
\label{met}
\end{eqnarray}
Equal time sections of this
metric are two-spheres, and there are no globally timelike Killing
vectors. 
However, there does exist a timelike Killing vector in certain patches of 
this spacetime. Figure 1 shows the Penrose diagram of 
global de Sitter space with these patches - II and IV. These regions  
are causally disconnected and the timelike Killing vector flows in opposite
directions in these two patches. Each of these patches is bounded by the
cosmological horizon, and described by the metric
\begin{eqnarray}
ds^2~=~ - N^2~dt^2 + N^{-2}~dr^2 + r^2~d\phi^{2}
\label{met2}
\end{eqnarray}
where
\begin{eqnarray}
N^{2}~=~(1~-~\frac{r^2}{l^2}),
\label{lapse}
\end{eqnarray}
and $0\leq r \leq l$. $\phi$ is an angular coordinate with
period $2\pi$. Since this metric is static, the patches II and IV are referred to as static patches.
The cosmological horizon in these coordinates is 
therefore at $r=l$. Constant $t$ surfaces are discs $D_{2}$,
and the topology of the patch is $D_{2} \otimes R$.
\begin{figure}[t]
\centerline{\epsfxsize 3in
           \epsfbox{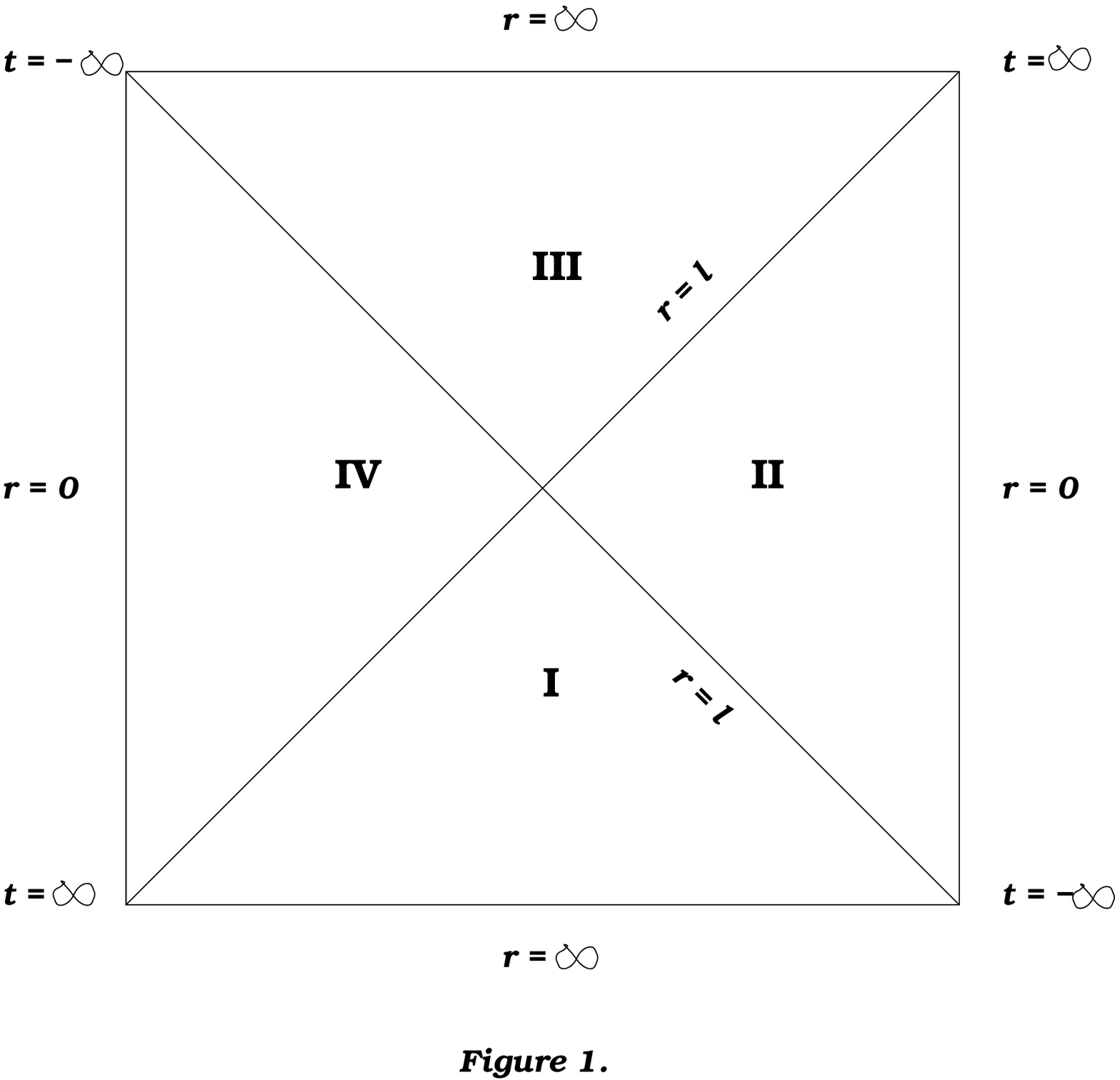}}
\end{figure}

Using (\ref{b8}), the connection $A^a$  corresponding
to the metric (\ref{met2}) may be written as:
\begin{eqnarray}
A^{0}& =& N (-d\phi + \frac{i}{l} dt) \nonumber \\
A^{1}& =& \frac{i}{l N} dr \nonumber \\
A^{2}& =& \frac{r}{l^2} dt + \frac{i r}{l} d\phi
\label{conn}
\end{eqnarray}

\section{Euclidean de Sitter space}
The Euclidean gravity action is the difference of two 
$SU(2)$ Chern-Simons actions - where the connections 
corresponding to the two actions are real and given by
$A ~=~ \left(\omega^a + \frac{1}{l}~ e^a\right) T_a $ and
$B ~=~ \left(\omega^a - \frac{1}{l}~ e^a\right) T_a$.

As is well-known, the static patch of de Sitter space (either
patches II or IV in Fig. 1) represents the 
spacetime completely accessible to the timelike observer - whose
observable universe is bounded by a cosmological horizon. Entropy of de Sitter
space is associated then with loss of information across the horizon. Thus,
the relevant space for a computation of de Sitter entropy is the static patch
of de Sitter space. Similar ideas about the relevance of the static patch (or the
"causal diamond") to de Sitter entropy were suggested in \cite{bousso}. In \cite{ms},
the entropy associated with de Sitter space was computed using the Lorentzian static
patch. This, as we saw, has the topology $D_{2} \otimes R $, (where $R$ is the 
time direction) and the degrees of freedom corresponding to the entropy in this
computation resided on the boundary cylinder.

We would like to compute de Sitter entropy in a Euclidean picture. As we shall see,
we reproduce the entropy as in previous Lorentzian computations - but also get
new results on the nature of quantum fluctuations in de Sitter space.
We therefore consider the Euclidean continuation of the metric
on the static patch (\ref{met2}).
This is obtained by taking $t_{E} = it$.
The metric is
\begin{eqnarray}
ds^2~=~  N^2~dt_{E}^{2} + N^{-2}~dr^2 + r^2~d\phi^{2}
\label{eucmet2}
\end{eqnarray}
where $N$ is given by (\ref{lapse}). In addition, for regularity of the metric,
we make 
the Euclidean time periodic with its period
$\beta = 2\pi l$.
Here,  $0 \leq r \leq l$. The periodicity of the Euclidean time now changes
the topology of each of the static patches II and IV 
from the Lorentzian $D_{2} \otimes R$ to $D_{2} \otimes S^1$ -
where the $S^1$ direction is the compactified Euclidean time. This is
nothing but a solid torus - and in the Euclidean picture, we expect the 
degrees of freedom corresponding to de Sitter entropy to be associated with
this solid torus.

Thus we are interested in the Euclidean gravity partition function 
studied through two $SU(2)$ Chern-Simons theories on a solid torus.
Corresponding to the metric (\ref{eucmet2}), 
the connections for the two $SU(2)$ Chern-Simons theories are given by: 
\begin{eqnarray}
A^{0} &=& - N (d\phi + \frac{1}{l} dt_{E}) \nonumber \\
A^{1} &=&  \frac{1}{l N} dr \nonumber \\
A^{2} &=& - \frac{r}{l^2} dt_{E} + \frac{r}{l} d\phi    
\label{aconn}
\end{eqnarray}

\begin{eqnarray}
B^{0} &=& N (-d\phi + \frac{1}{l} dt_{E}) \nonumber \\
B^{1} &=& - \frac{1}{l N} dr \nonumber \\
B^{2} &=& - (\frac{r}{l^2} dt_{E} + \frac{r}{l} d\phi) 
\label{bconn}
\end{eqnarray}

\section{Partition function for quantum gravity in de Sitter space}

We use the Chern-Simons formulation of gravity to construct a 
partition function for gravity in de Sitter space. 
We saw in the last section that the manifold of interest 
is a solid torus. This is also the topology of the Euclidean
BTZ black hole. This enables us to use the construction developed in
\cite{gks} for Euclidean BTZ black holes here too.

In order to compute the de Sitter partition function, we first evaluate
the Chern-Simons path integral on a solid torus. This path integral has been
discussed in \cite{elit}, \cite{lab1}, \cite{lab2} and \cite{hayashi}.
Through a suitable  gauge transformation, the connection is set to a 
constant value on the toroidal boundary. 
In terms of coordinates on the toroidal boundary $x$ and $y$ with unit period,
we can define $z = (x + \tau y)$ such that 
\begin{eqnarray}
\int_{a} dz~ =~ 1,~~~~~ \int_{b} dz~ =~ \tau
\label{cycle}
\end{eqnarray}
where $a$ is the contractible cycle and $b$ the non-contractible cycle of 
the solid torus and $\tau=\tau_1+i\tau_2$ is the modular parameter of
the boundary torus.
Then, the connection can be written as \cite{lab1}:
\begin{eqnarray}
A ~=~ \left(\frac{-i \pi \tilde u}{ \tau_2}~ d\bar z + \frac{i \pi u}{ \tau_2} 
~dz\right) T_3
\label{adefn}
\end{eqnarray}
where $u$ and $\tilde u$ are canonically conjugate fields and obey 
the canonical commutation relation:
\begin{eqnarray}
[\tilde u, u]~ =~ \frac{2 \tau_2}{\pi (k+2)}
\label{ccr}
\end{eqnarray}
Since $A$ is an $SU(2)$ connection, $\tilde u = \bar u$, 
where $\bar u$ is the complex conjugate of $u$.
For the case of the BTZ black hole, the information about the black hole
parameters $M$ and $J$ is contained in the non-trivial holonomy of the connection $A$ 
around the non-contractible cycle $b$. 
In the de Sitter case the Euclidean time is associated with the non-contractible
$b$ cycle of the solid torus. This is so because the Euclidean continuation of the
Lorentzian static patch $D_{2} \otimes R$, with $R$ as world line of the
timelike observer sitting at the center of the disc involves, besides a Wick rotation,
a compactification of this line to obtain a solid torus.
The holonomy of the connection around the non-contractible cycle is
therefore related to the period $\beta$ of the Euclidean time. Similarly,
the holonomy around the contractible cycle is related to the
periodicity of the $\phi$ coordinate.  Traces
of the holonomies around the contractible $a$ cycle and 
non-contractible $b$ cycle from connections in (\ref{aconn}) are:
\begin{eqnarray}
Tr(H_a)~ =~ 2 \cos (\Theta), ~~~~~Tr(H_b)~ =~ 
2 \cos \left(\frac{\beta}{l} 
\right)
\label{trh}
\end{eqnarray}
Classical de Sitter space corresponds to a value of $\Theta = 2\pi$. 
Now in order that 
the gauge fields (\ref{adefn}) exhibit the same holonomies 
as above, we have                                                          
\begin{eqnarray}
A_{z}~ =~ \frac{-i \pi}{\tau_2}~ \tilde u, ~~~~~~~ 
A_{\bar z}~ = ~\frac{i \pi}{\tau_2}~ u
\end{eqnarray}
with 
\begin{eqnarray}
u ~=~ \frac{1}{2 \pi} \left(\Theta \tau - \frac{ 
\beta}{l}\right),
~~~~~
\tilde u ~=~ \frac{1}{2 \pi} \left(\Theta \bar \tau - \frac{
\beta}{l}\right)
\label{uubar}
\end{eqnarray}

Next we write the 
Chern-Simons path integral on a solid torus with a
 boundary modular parameter $\tau$. For
 a fixed boundary value of the connection, i.e. a fixed value of $u$, this path integral is formally
 equivalent to a state $\psi_{0}(u, \tau)$ with no Wilson lines in the solid torus.
 The states corresponding  to having closed Wilson lines (along the
 non-contractible cycle) carrying spin $j/2$
 ($j \le k$) representations in the solid torus
 are given by \cite{elit}, \cite{lab1}, \cite{lab2},
 \cite{hayashi}:
 \begin{eqnarray}
 \psi_{j}(u, \tau)~ =~ \exp\left\{ {\pi k\over4\tau_2}\,u^2 \right\}
 ~\chi_{j}(u, \tau) ,
 \label{c3}
 \end{eqnarray}
 where $\chi_{j} $ are the Weyl-Kac characters for affine $\hbox{SU}(2)$
 which can be expressed in terms of the well-known
 Theta functions as
 \begin{eqnarray}
 \chi_{j}(u, \tau)~ =~ \frac{\Theta_{j +1}^{(k+2)}(u, \tau, 0)~-~\Theta_{-j -1}^{(k+2)}(u, \tau, 0)}
 {\Theta_{1}^{2}(u, \tau, 0)~-~\Theta_{-1}^{2}(u, \tau, 0)}
 \label{c4}
 \end{eqnarray}
 where Theta functions are given by:
 \begin{eqnarray}
 \Theta_{\mu}^{k}(u, \tau, z) ~=~
 \exp (-2 \pi i k z)~ \sum_{n \in \cal Z} \exp 2 \pi i k \left[(n + \frac{\mu}
 {2 k})^2 \tau ~+~(n + \frac{\mu}{2 k}) u \right]
 \label{theta}
 \end{eqnarray}
  
As in the computation in \cite{gks} for the BTZ black hole, the 
de Sitter partition function is constructed from the boundary state
$\psi_{0}(u, \tau)$. The construction is motivated by the
following observations :

(a) In the Chern-Simons functional integral over a solid torus, we shall 
integrate over all gauge connections with fixed holonomy $H_b$ around
the non-contractible cycle. This corresponds to the partition function
with fixed period $\beta$ of the Euclidean time, that is, fixed inverse
temperature. This in 
turn means we are dealing with the canonical ensemble.
The variable conjugate to this holonomy is the holonomy
around the other (contractible) cycle, which is {\em not} fixed
any more to the classical value given by $\Theta = 2\pi$ for
de Sitter space. We must sum over contributions from all possible
values of $\Theta$ in the partition function.
This corresponds to starting with the value of $u$ for the classical solution,
i.e. with $\Theta = 2\pi$
in (\ref{trh}), and then considering all other shifts of $u$ of the form
\begin{eqnarray}
u ~\rightarrow~ u + \alpha \tau
\end{eqnarray}
where $\alpha$ is an arbitrary number. This is implemented by a translation operator of the form
\begin{eqnarray}
T ~=~ \exp \left(\alpha \tau \frac{\partial}{\partial u}\right)
\end{eqnarray}
However, this operator is not gauge invariant. The only gauge-invariant way of implementing these
translations is through Verlinde operators of the form
\begin{eqnarray}
W_{j}~ = ~\sum_{n \in \Lambda_{j}} \exp \left(\frac{-n \pi \bar \tau u}{\tau_{2}} +
\frac{n \tau}{k+2} \frac{\partial}{\partial u} \right)
\end{eqnarray}
where $\Lambda_{j} ~=~ {-j, -j+2,...,j-2, j}$.
This means that all possible shifts in $u$ are not allowed. The
only possible shifts allowed by gauge invariance are:
\begin{eqnarray}
u ~\rightarrow ~u + \frac{n \tau}{k+2}
\label{ushift}
\end{eqnarray}
where $n$ is always an integer taking a maximum value of $k$.
Thus, the only allowed values of $\Theta$ are $2\pi(1 + \frac{n}{k+2})$.
We know that acting on the state with no Wilson lines in the solid torus with the
Verlinde operator $W_{j}$
corresponds
to inserting a Wilson line of spin $j/2$ around the non-contractible cycle.
Thus, taking into account all states with different
shifted values of $u$ as in $(\ref{ushift})$
means that we have to take into
account all the states  in the boundary corresponding to the insertion 
of such Wilson lines. These are the
states $\psi_{j}(u, \tau)$ given in (\ref{c3}).
 
(b) In order to obtain the final partition function, we must also integrate over all
values of the modular parameter $\tau$, i.e. over all inequivalent tori 
with the same holonomy around the non-contractible cycle.
The integrand, which is a function of $u$ and $\tau$, 
must be the square of the partition function of a
gauged $SU(2)_{k}$ Wess-Zumino model  corresponding to the two $SU(2)$
Chern-Simons theories.
It must be modular invariant -- modular invariance corresponds to 
invariance under large diffeomorphisms
of the torus. 
The partition function is then of the form
\begin{eqnarray}
Z ~= ~ \int d\mu(\tau, \bar \tau) ~\left|~\sum_{j=0}^{k} ~a_{j}(\tau)~ \psi_{j}
(u, \tau)~\right|^2
\label{pf}
\end{eqnarray}
where $d\mu(\tau, \bar \tau)~ =~ \frac{d\tau d\bar \tau}{\tau_{2}^{2}}$ is the 
modular invariant measure, and the integration is over a 
fundamental domain in the $\tau$ plane. 
Coefficients $a_{j}(\tau)$ must be chosen  such that the 
integrand is modular invariant.

As discussed in  \cite{gks},  these coefficients are given by 
$a_j(\tau)~=~(\psi_j(0,\tau))^*$ so that the partition function is uniquely
fixed to be
\begin{eqnarray}
Z_{dS} = \int d\mu(\tau, \bar \tau) \left|~\sum_{j=0}^{k}
~(\psi_{j}(0, \tau))^{*} ~\psi_{j}(u, \tau)~\right|^2
\label{dspf}
\end{eqnarray}
This is an {\em exact} expression for the canonical partition function
of quantum gravity on $dS_{3}$. We now proceed to compute the partition
function by substituting in the expression above the values of 
$u$ and $\bar u$ from (\ref{uubar}) with $\Theta = 2\pi$. We work
in the regime where $k$ (and therefore $l$) is large. Also, we must
perform an analytic continuation to get the Lorentzian result - this is
done by taking $G \rightarrow -G$, and $\beta \rightarrow i\beta$.
For the regime when $k$ is large, the leading contribution to the
sum in the integrand comes from $j=0$ as in \cite{gks}. The $\tau_{2}$
integral can in fact be done exactly. 
We have
\begin{eqnarray}
Z_{dS} = \int_{-1/2}^{1/2} d\tau_{1}~ 4\pi~~ e^{\beta~ k/2l}~~ \frac{1}{f(\tau_{1})}~~ K_{1}(-k/2~~ f(\tau_{1}))
\label{tau1int}
\end{eqnarray}
where $f(\tau_{1}) = \sqrt{ \frac{\beta^2}{l^2} - 4\pi^2 \tau_{1}^{2}}$, and $K_{1}$ is the Bessel function
of imaginary argument.
Using the approximation for the Bessel function with large argument
\begin{eqnarray}
K_{1}(z) = \sqrt{\frac{\pi}{2 z}} e^{-z} [1 + O(\frac{1}{z}) +...]
\label{asyk}
\end{eqnarray}
with the replacement $\beta = 2\pi l$ for de Sitter space, we get, in the large $k$ regime :

\begin{eqnarray}
Z_{dS} = 4 \sqrt{\pi}~~ \frac{4G}{2\pi l}~~ e^{2\pi l/4 G}
\label{cpf}
\end{eqnarray}

The form of the partition function indicates that at leading order, 
it is of the form $e^{S}$,
where $S~=~\frac{2\pi l}{4G}$ is the semi-classical entropy. Since this 
is the partition function in the canonical
ensemble, we would have expected an additional term $e^{-i\beta E}$
where $E$ is the energy of
de Sitter space. The notion of energy in asymptotically de Sitter spaces needs to be defined carefully,
due to the absence of a global timelike Killing vector.
The energy $E$ that emerges in our formalism is defined on the horizon, 
and not at asymptotic
infinity, as has been done, for e.g in \cite{vb}. 
Our result seems to indicate that that energy
$E$ is zero for de Sitter space.
Such a result coincides with the definition of energy
as given by Abbott and Deser \cite{AD}. 
The canonical partition function at leading order is 
therefore the same as the density
of states. The multiplicative prefactor in (\ref{cpf}) is the leading correction to the semi-classical
result. The entropy is therefore
\begin{eqnarray}
S = (2\pi l)/4G ~-~\log \frac{2\pi l}{4G} + ........
\label{entropy}
\end{eqnarray}
The leading term is the semi-classical Bekenstein-Hawking entropy
that is proportional to the horizon ``area". The second term is the
leading correction that is logarithmic in area.
In the following sections, we discuss in detail the results 
we have obtained - on the nature of the quantum fluctuations,
and the logarithmic correction we have seen above.

\section{The nature of the quantum fluctuations}
In our set-up, from the choice of ensemble and considerations of gauge
invariance, the partition function (\ref{dspf}) involved insertion
of closed Wilson lines of spins $j/2 \neq 0$, $j \le k$. These 
correspond to defects centered at the origin of the $\phi$ coordinate
in (\ref{eucmet2}), i.e around the worldline of the timelike observer.
As is well-known, such defects correspond to point masses in 
$dS_{3}$ \cite{deser}. A point mass in $dS_{3}$ can be described in static
coordinates by the metric
\begin{eqnarray}
ds^2~=~ - N^2~dt^2 + N^{-2}~dr^2 + r^2~d\phi^{2}
\label{pointmass}
\end{eqnarray}
where now
\begin{eqnarray}
N^{2}~=~(8GM~-~\frac{r^2}{l^2}),
\label{masslapse}
\end{eqnarray}
and $0\leq r \leq r_{+}$ with $r_{+}~=~l\sqrt{8GM}$ as the radius of the
cosmological horizon. 
Let us consider the static patch metric (\ref{pointmass}) of $dS_{3}$. Near $r=0$, the
origin of the $\phi$ coordinate, it looks like Minkowski space. 
Then metric (\ref{pointmass}) near $r=0$  can be rewritten as the Minkowski metric
\begin{eqnarray}
ds^2 = -dt_{1}^{2} + dr_{1}^{2} + r_{1}^{2} d\phi_{1}^{2}
\end{eqnarray}
where $t_{1} = \sqrt{8GM} t$, $r_{1} = r/\sqrt{8GM}$ and
$\phi_{1} = \sqrt{8GM} \phi$.
But since $\phi$ had a periodicity $2\pi$, $\phi_{1}$ has a 
periodicity $2\pi \sqrt{8GM}$. Thus the deficit angle at 
$r=0$ is $2\pi (1 - \sqrt{8GM})$, i.e $2\pi (1 - \frac{r_{+}}{l})$.
When $r_{+}=l$, the deficit is zero, and we have de Sitter space.

In the partition function (\ref{dspf}), there is a discrete sum
over deficits in the $\phi$ coordinate - from (\ref{ushift}), the
deficits are of the form $2\pi~ \frac{n}{k+2}$ where $n$ is 
an integer taking the maximum value of $k$. $n=0$ corresponds to 
de Sitter space, i.e no deficit. Thus from the discussion above, 
we see that $(1 - \frac{r_{+}}{l}) = \frac{n}{k+2}$. The maximum
deficit possible is  $n=k$. Thus
there seems to be a maximum allowed value for the "mass" of the
point particle, $M$. Interestingly, this relation also
implies that $r_{+}$, and therefore the mass of the point particle
takes only a {\em discrete} number of values, 
labelled by the integer $n$.
Though classically all values of deficit angle between
$0$ and $2\pi$ are allowed, gauge invariant quantization allowed
only discrete set of values.   

There is another interesting observation :
In the computation of the de Sitter partition function (\ref{dspf}),
we set $\Theta=2\pi$ - which corresponds to de Sitter space. Thus
the partition function describes quantum fluctuations on the de
Sitter background. We could instead choose any value of $\Theta$
from $0$ to $2\pi$. Then the partition function 
would correspond to fluctuations around a background with the
corresponding point mass. What does that partition function look
like? The exercise done in the previous section can be repeated for
this case - and surprisingly, the leading answer is the same! Quantum
fluctuations of any such background are point masses taking the same
set of discrete values as above. The dominant contribution
to the partition function again comes from spin $j=0$, which 
corresponds to {\it empty} de Sitter space.  Doing the computation 
carefully (taking into account the changed value of the temperature),
we find that to the leading order, the entropy is the 
same as that of de Sitter space, i.e $S = (2\pi l)/4G$.
However, the logarithmic corrections to the entropy are more
complicated now and carry the information about the point mass parameter. 
All this strongly suggests
a quantization of point mass configurations in a quantum 
theory of gravity on $dS_{3}$. Also, since the leading contribution 
to the entropy always comes from empty de Sitter space, this 
presents an explicit realisation of the entropy bound of Bousso
\cite{bousso} in three dimensions.

\section{The log(area) correction to the semi-classical entropy}
We note that in the expression for entropy (\ref{entropy}),
the correction to the Bekenstein-Hawking entropy is logarithmic
in area, with a coefficient $-1$. The logarithmic correction has
been observed in many computations of black hole entropy in 
quantum gravity. It was
first computed for the (3+1)-d Schwarzschild black hole in 
the quantum geometry formulation of gravity - where,
for a large massive black hole,  
the next order log(area) correction
had a numerical coefficient, $-3/2$ \cite{majumdar}.
Subsequently, this correction (with the same numerical
coefficient!) has been seen in computations of the (2+1)-d BTZ black hole
entropy in various approaches \cite{majumdar}, \cite{gks}, 
leading to the question of
whether this coefficient is universal. Below, we clarify
several issues related to the universality of this coefficient. 
Incidentally, the logarithmic
correction with a different coefficient was seen in a one-loop 
computation of the correction
to the entropy of the BTZ black hole (of small horizon area) 
due to a scalar field \cite{sm}. 
However, our discussion of the logarithmic coefficient is 
the correction due to quantum
gravity fluctuations, and distinct from corrections due 
to scalar fields or other matter
coupled to the black hole. 

The computation of BTZ black hole entropy in \cite{gks} was
done in the same (Chern-Simons) formulation as the de Sitter
case, and the numerical coefficient of the logarithmic term was
$-3/2$, whereas for the de Sitter case, it is $-1$.  
This is somewhat puzzling at first glance. The 
black hole entropy was computed in the regime $r_{+} >> l$, where
$r_{+}$ is the 
black hole horizon radius and $l$ is the $AdS$ radius of 
curvature. Then, there was an integral over the modular 
parameter similar to (\ref{dspf}). The saddle-point for $\tau_{2}$, 
the imaginary part of the modular parameter occured when $\tau_{2} = 
r_{+}/l$. Thus this was the regime when $\tau_{2}$ was large. 
An interesting 
observation was made in \cite{gks} that replacing $r_{+}/l$
in the black hole partition function  
by $l/r_{+}$, where now $r_{+} << l$, the $AdS$ gas partition function
was obtained, with the coefficient of the correction being $+3/2$. 
This corresponds to a situation where the modular parameter 
$\tau_{2} = r_{+}/l$ is small. What happens when $r_{+} \sim l$, 
i.e $\tau_{2} \sim 1$? In fact, this is very similar to the de Sitter
case, since the de Sitter horizon radius is exactly $l$! The computation
follows similar lines and leads to similar results. It can in fact
be verified directly from (\ref{dspf}) that the saddle-point is at 
$\tau_{2} = 1$. Here, we see that the coefficient of the logarithmic
correction is $-1$. Thus, the coefficient of the correction seems to depend
on the regime one is looking at. When, as in the above case, there are two
independent length parameters $l$ and $r_{+}$, only for $r_{+} >> l$ do we
get the coefficient $-3/2$. 

\noindent Summarising our result for BTZ black hole we find:
\begin{eqnarray}
~For~~r_+ >>~l \qquad S~&=&~\frac{2\pi r_+}{4G} ~-~\frac{3}{2}~
\log~( \frac{2\pi r_+}{4G}) ~+~\cdots \nonumber \\
~~~~~~r_+ =~l \qquad S~&=&~\frac{2\pi r_+}{4G} ~-~
\log~(\frac{2\pi r_+}{4G})~+~\cdots \nonumber \\                 
~~~~~~r_+ <<~l \qquad S~&=&~\frac{2\pi l^2}{4r_+ G} 
~+~\frac{3}{2}~\log(\frac{r_+}{l}) ~+~\cdots
\label{regimetb}
\end{eqnarray}
where the last expression in (\ref{regimetb}) for $r_{+} <<~l$ is the 
entropy of the AdS gas.

The above results are reminiscent of a duality proposed in \cite{corichi} 
between the Euclidean BTZ black hole and Lorentzian de Sitter spaces. There,
the classical holonomy of the connection in the black hole case was related
to the ratio $r_{+}/l$. From the duality, 
this was also the holonomy of the connection 
in a de Sitter space with a point mass, the mass being related to the parameter $r_{+}$.
In particular, vacuum de Sitter space corresponds to $r_{+} = l$. Although this
duality is only at the level of {\em actions} (for a Lorentzian theory with positive $\Lambda$
and a Euclidean one with negative $\Lambda$), we find here that a BTZ black hole with
horizon radius $r_{+} = l$ and vacuum de Sitter space have the same entropy - both
at the leading and sub-leading order!

Let us examine (\ref{tau1int}) closer to understand how the 
coefficient of the logarithmic
correction in the de Sitter case is $-1$. 
Using the asymptotic expansion of the Bessel
function (\ref{asyk}), we see that the $\tau_{2}$ 
integration contributes a logarithmic
term with a coefficient $-1/2$. The $\tau_{1}$ integration 
also contributes the same and the coefficient is thus $-1$. 

Entropy of de Sitter space can also studied  from an 
alternative point of view by using dS/CFT correspondence \cite{strominger}.
In this framework all the information about quantum gravity in the bulk is
expected to be contained in the conformal field theory at 
past or future infinity. 
The CFT is described by considering all possible metric fluctuations keeping the
asymptotic behaviour to be de Sitter space. It consists of two copies
of Virasaro algebras, each with central charge 
$c = 3l/2G$. As shown in \cite{park, vb, myung}, the 
eigenvalues of the Virasoro generators 
$L_{0}$ and $\bar L_{0}$ for de Sitter space are
both equal to $l/8G$. Using the Rademacher expansion for modular forms, one can 
generalize the Cardy formula for growth of states in a CFT beyond the leading term.
In \cite{bsen}, the sub-leading correction to the entropy
of a BTZ black hole was determined from this generalisation. 
We use these results to find the sub-leading
corrections to the de Sitter entropy from the dS/CFT correspondence. From 
\cite{bsen}, the entropy obtained from a CFT with a
given the central charge $c$ and eigenvalue of the
Virasoro generator $L_0~=~N$, is given by 
\begin{eqnarray}
S_1 = S_{0} - 3/2 \log S_{0} + \log c + .......
\label{entr}
\end{eqnarray}
where $S_{0} = 2\pi \sqrt{\frac{c}{6}(N - \frac{c}{24})}$.
This is the contribution from the Virasoro generator $L_{0}$. There is a similar
contribution $S_2$ associated with the Virasoro generator 
$\bar L_{0}$, given by replacing $N$
in the above by $\bar N$, the eigenvalue of $\bar L_{0}$. 

Substituting
$c = 3l/2G$ and $N = \bar N = l/8G$ in the above, we see that 
\begin{eqnarray}
S = S_1 + S_2 = 2\pi l/4G - \log \frac{2\pi l}{4G} + ....
\label{cftentr}
\end{eqnarray}
with the same coefficient $-1$ for the logarithmic correction as that obtained
from the gravity partition function (\ref{cpf}) in (\ref{entropy}). Here, the
contribution from each of $S_1$ and $S_2$ to the logarithmic correction was 
$-1/2 ~\log \frac{2\pi l}{4G}$. 

Thus, the quantum gravity calculation of de Sitter entropy and the 
entropy computation from the asymptotic CFT agree even in the sub-leading
correction to the Bekenstein-Hawking term. 

\section{Acknowledgement}

$V.~S$ would like to acknowledge support of a fellowship from the 
Alexander von Humboldt Foundation.

\end{document}